\documentclass[11pt]{article}

\usepackage[final]{acl}

\usepackage{times}
\usepackage{latexsym}

\usepackage[T1]{fontenc}

\usepackage[utf8]{inputenc}

\usepackage{microtype}

\usepackage{inconsolata}

\usepackage{graphicx}

\usepackage{tcolorbox}

\usepackage{amsfonts}
\usepackage{amsmath}
\usepackage{makecell}

\usepackage{booktabs}

\usepackage{pifont}
\usepackage{xcolor}
\newcommand{\cmark}{\textcolor{green}{\ding{51}}}
\newcommand{\xmark}{\textcolor{red}{\ding{55}}}

\title{FIGMA: Towards FIne-Grained Music retrievAl}

\author{
  \textbf{Nishit Anand\textsuperscript{1}},
  \textbf{Ashish Seth\textsuperscript{1}},
  \textbf{Sreyan Ghosh\textsuperscript{1}},
  \textbf{Dinesh Manocha\textsuperscript{1}}\thanks{\ Equal Advising.},
  \textbf{Ramani Duraiswami\textsuperscript{1}}\footnotemark[\value{footnote}]
\\
\\
  \textsuperscript{1}University of Maryland, College Park, USA
\\
  \textbf{Correspondence:} \href{mailto:nishit@umd.edu}{nishit@umd.edu} \\
  \centering Project: \url{https://nishitanand.github.io/figma-website}
}

\begin{document}
\maketitle
\begin{abstract}
Retrieving music using natural language descriptions has improved with contrastive audio–text models such as CLAP, but current systems remain limited to coarse semantic queries. When descriptions specify fine-grained musical attributes such as tempo, key, chord progression, or rhythmic structure, existing models often fail to retrieve the correct audio. We show that this limitation stems from the contrastive learning objective itself: despite being trained on long captions, CLAP-based models effectively utilize only the first few tokens, discarding much of the information encoded in detailed prompts. Then, we propose \textbf{FIGMA} (\textbf{FI}ne-\textbf{G}rained \textbf{M}usic Retriev\textbf{A}l), a multi-view contrastive architecture that addresses this limitation by jointly optimizing global audio–text alignment and frame-level, token-wise alignment. This design enables FIGMA to capture both high-level semantic context and fine-grained musical attributes within a unified representation space. Moreover, we formalize the task of Fine-Grained Music Retrieval and construct Fine-Grained Music Caption dataset (FGMCaps), a large-scale dataset of 380K music–caption pairs for training along with a 10K test set, both annotated with tempo, key, chord progression, beat count, as well as genre and mood. Extensive experiments demonstrate that FIGMA consistently outperforms existing CLAP-based music retrieval models across multiple music retrieval benchmarks, including out-of-domain evaluations, with relative improvements of up to 73.3\%.
\end{abstract}

\begin{figure}[t]
  \centering
  \includegraphics[width=\columnwidth]{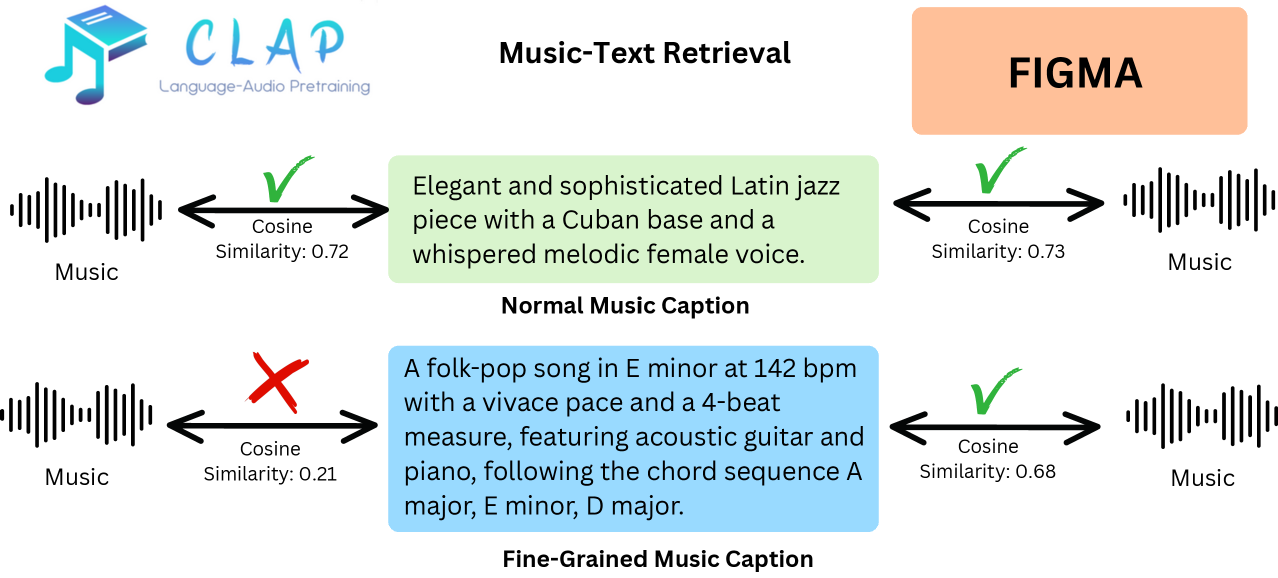}
  \caption{Current retrieval models struggle to retrieve music with fine-grained captions, whereas our proposed method, FIGMA is able to understand dense fine-grained captions to retrieve music.}
  \label{fig:hero_diagram}
\end{figure}

\section{Introduction}

\begin{figure*}[t]
  \centering
  \includegraphics[width=\textwidth]{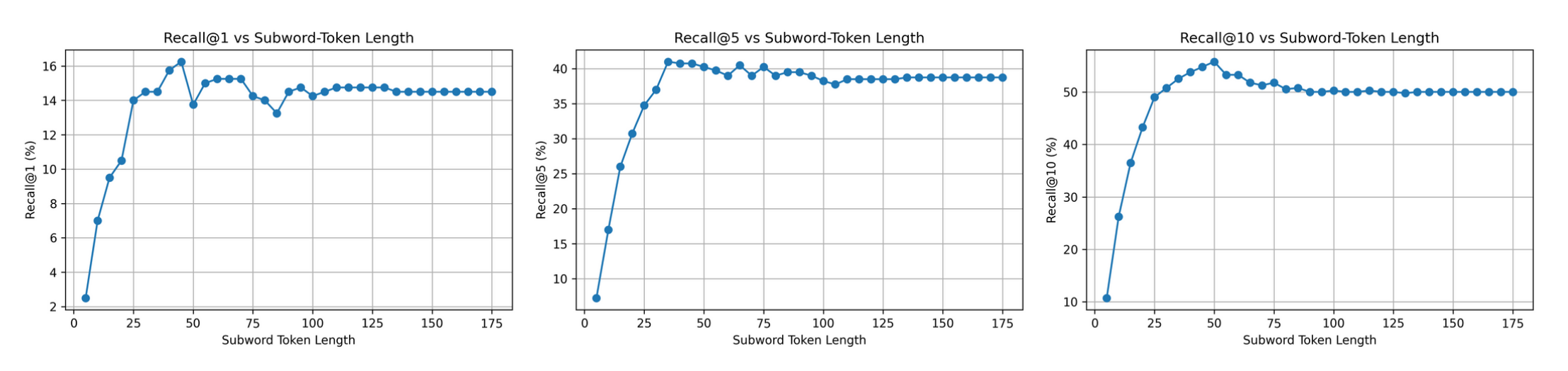}
  \caption{Retrieval @1{,} 5, and 10 performance of MuQMuLaN on the MusicBench Test Set. We observe that retrieval performance saturates beyond 50 tokens, indicating that current retrieval models are unable to fully understand and use detailed captions.}
  \label{fig:token_length_expt}
\end{figure*}

Music retrieval refers to the automatic matching of natural language descriptions to relevant audio tracks within large music collections. This capability is used in a wide range of applications, including creating personalized recommendation engines, on-the-fly playlist curation, and rapid content discovery and matching for both casual listeners and professional musicians. Early music retrieval systems relied on handcrafted audio descriptors, such as MFCCs \cite{mfcc1, logan2000mfcc} and chroma vectors \cite{chroma-vectors-1, chroma-vectors-2}, and measured similarity using fixed distance metrics or simple metric-learning schemes. Modern approaches instead embed text queries and audio clips into a shared representation space and optimize contrastive objectives. Following the success of CLIP in computer vision \cite{clip}, this paradigm was extended to audio through models such as CLAP \cite{laion-clap}. While general-purpose audio--text models, including LAION CLAP \cite{laion-clap} and Microsoft CLAP \cite{msclap22, msclap23}, achieve strong zero-shot performance across diverse audio tasks, they often underperform on music-specific retrieval.

\noindent To address this gap, prior work has proposed contrastive architectures tailored to music, including MuLaN \cite{mulan}, CLAMP \cite{clamp1}, and MuQMuLaN \cite{muqmulan}. These models are trained exclusively on music--text pairs and have achieved substantial gains in music retrieval and music understanding. However, despite their effectiveness, they struggle with detailed textual queries that specify precise musical attributes, posing a significant challenge in practical settings. For example, a song composer who wants to search a music clip from a personal collection of 10K--20K tracks would use a query such as ``a song in F major at 110 BPM with a 4/4 beat pattern.'' Rather than relying on coarse descriptors like genre or mood, such queries require retrieval based on chord structure, tempo, key, and rhythmic profile---criteria under which existing models often fail to return appropriate matches.

\noindent We refer to this task of retrieving music clips using detailed textual descriptions that include both high-level attributes (e.g., genre, instrumentation) and precise musical parameters (e.g., key, chord progression, tempo, beat count) as \textit{Fine-Grained Music Retrieval}. This capability is essential for enabling musicians and producers to efficiently locate and repurpose audio material according to exact musical specifications.

\noindent We show that existing contrastive audio--text models fail to effectively utilize long, detailed music captions. Empirically, retrieval performance saturates once captions exceed 40--50 tokens, indicating that additional musical information in longer descriptions is largely ignored. We demonstrate this behavior on the MusicBench test set \cite{mustango}, where models consistently underperform on queries with fine-grained attributes such as tempo, chord progression, and beat count. Furthermore, fine-tuning LAION CLAP on the FGMCaps training split yields only marginal improvements despite the presence of richly detailed captions.

\noindent This limitation stems from a fundamental property of standard contrastive learning objectives. Existing architectures collapse both audio and text into single global embeddings; typically by mean-pooling audio features over time and representing captions with a single CLS token. This global aggregation discards temporal structure in the audio and token-level distinctions in the text, causing long captions to behave as bags of words. Consequently, even when fine-grained musical attributes are present in captions, these models lack a mechanism to align them with corresponding features in the audio.

\noindent To address this challenge, we propose FIGMA, a novel architecture incorporating a Multi-View Contrastive loss. In addition to a standard global contrastive objective, we introduce a frame-level, token-wise contrastive loss that explicitly aligns audio frames with caption tokens. This dual-level training objective preserves both global semantic features and fine-grained correspondence between musical attributes and their textual descriptions, leading to substantially improved retrieval performance. We summarize our contributions as follows:

\begin{enumerate}
\item We identify a fundamental limitation of existing CLAP-based models for music retrieval, showing that standard contrastive objectives cause long captions to collapse into coarse representations, with tokens beyond the first 40–50 contributing little to retrieval performance. To address this, we formalize the task of \textbf{Fine-Grained Music Retrieval} and propose \textbf{FIGMA}, a novel architecture based on a Multi-View Contrastive loss that jointly optimizes global audio–text alignment and frame-level, token-wise alignment.
\item We introduce \textbf{FGMCaps}, a large-scale dataset for fine-grained music retrieval, comprising 380K music–caption pairs for training and a 10K test set annotated with music-centric attributes including tempo, key, chord progression, beat count, genre and mood metadata.
\item Through extensive experiments, we demonstrate that FIGMA consistently outperforms existing CLAP-based music retrieval models across multiple benchmarks, including out-of-domain evaluations, with relative improvements of up to 73.3\%.
\end{enumerate}

\section{Related Work}

Contrastive learning for classification and retrieval first gained attention with the introduction of CLIP \cite{clip} in the computer vision domain. In the audio research field, CLAP (Contrastive Language–Audio Pretraining) \cite{laion-clap} has similarly demonstrated impressive zero-shot audio and music retrieval capabilities. Early general-purpose audio models adopted contrastive objectives, most notably LAION-CLAP \cite{laion-clap}, trained on 630 K audio–text pairs, and Microsoft’s CLAP 2022 \cite{msclap22} and its 2023 successor \cite{msclap23}, the latter trained on 4.6 million audio-text pairs. While these models excel at broad audio classification and retrieval, they often struggle with music-specific retrieval, which requires the model to have knowledge about musical features.

To address this gap, researchers have developed models trained on music–text datasets, typically music captions, using contrastive learning to achieve stronger music retrieval performance. The MuLaN model \cite{mulan}, for example, was trained on 44 million music recordings paired with textual annotations and exhibits robust music understanding and retrieval accuracy. More recent efforts continue to tailor CLAP-style architectures to the music domain: CLAMP \cite{clamp1} integrates sheet-music representations, whereas the MuQ paper \cite{muqmulan} introduces a dedicated music encoder and its MuQ-MuLaN variant, which processes raw audio directly in the style of LAION-CLAP and Microsoft-CLAP.

A concurrent line of work, FLAM \cite{flam}, also combines global and frame-level contrastive alignment for audio-language modeling. FIGMA differs from FLAM in two key respects. First, we adopt an InfoNCE-style frame-level objective rather than FLAM's SigLIP-style Binary Cross-Entropy (BCE) loss. BCE requires careful initialization of a logit bias $\beta$ to compensate for the severe 1:(B{-}1) positive-to-negative class imbalance, whereas InfoNCE sidesteps this issue through implicit softmax normalization, yielding more stable and hyperparameter-robust training. Second, rather than performing full model pre-training, we train only lightweight projection heads ($\sim$22M parameters) atop frozen MuQ and E5 encoders, making FIGMA substantially more compute-efficient while still achieving effective frame-level alignment.

\section{Methodology}

\noindent \textbf{CLAP Models Struggle with Long Captions.} To assess how effectively CLAP models leverage the full semantic content of long captions, we designed an experiment in which each text prompt was truncated to its first $k$ tokens, with $k$ varying from 5 up to the caption’s natural maximum in increments of 5. For a given $k$, every caption is shortened to its first $k$ tokens before retrieval, and we evaluate on Mustango's MusicBench Test set \cite{mustango}. At each truncation level, we record Retrieval@1, Retrieval@5, and Retrieval@10 to quantify how retrieval performance evolves as more tokens become available.

\noindent As shown in Figure~\ref{fig:token_length_expt}, all models’ Retrieval@1, @5, and @10 curves rise steeply at low $k$ but then plateau once captions exceed roughly 40-50 tokens. Beyond this threshold, adding further tokens yields negligible gains, indicating that the models fail to exploit the richer, fine-grained information encoded in longer prompts. We hypothesize that full utilization of every caption token - thereby capturing subtler musical attributes and descriptive nuances - would strengthen the alignment between audio clips and their textual descriptions, ultimately boosting downstream retrieval metrics.

\noindent We attribute this limited improvement to the inherent shortcomings of the vanilla contrastive objective employed by CLAP. In its standard form, the audio encoder’s output is mean-pooled across time to yield a single $d$-dimensional vector, while the text encoder uses the \texttt{[CLS]} token as a global summary of the caption. The contrastive loss is then applied only between these two pooled representations. By averaging over the audio frames and collapsing the token sequence into one embedding, frame-level acoustic nuances and token-level semantic details are discarded. Consequently, the model optimizes only for alignment of these coarse global features and cannot capture the richer, fine-grained correspondences that our detailed captions provide.

\begin{figure*}[t]
  \centering
  \includegraphics[width=\textwidth]{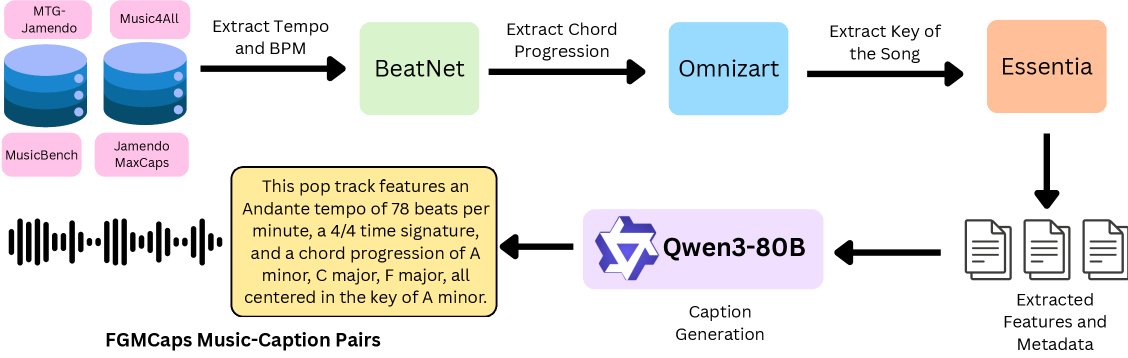}
  \caption{\small Our dataset construction pipeline consists of three stages. First, we do audio collection and preprocessing. Second, parallel automatic feature extraction applies BeatNet for tempo and beat count, Omnizart for chord progression, and Essentia KeyExtractor for musical key. Third, caption generation constructs prompts from extracted features and metadata, randomizes their order, and employs Qwen3-Next-80B-A3B-Instruct to generate coherent single-sentence descriptions. Quality control filters failed extractions and validates completeness, yielding 380K training, 10K validation, and 10K test audio-caption pairs.}
  \label{fig:dataset_construction}
\end{figure*}

\subsection{FIGMA}

\subsubsection{Overview}

For our architecture we use two frozen encoders: the MuQ audio encoder \cite{muqmulan}, pre-trained in a self-supervised manner on large-scale music data, and Microsoft’s Multilingual E5 Large Instruct text encoder \cite{e5}, selected for its robust language understanding.  The audio encoder receives a 10-second audio clip \(A\) sampled at 24kHz and the text encoder receives the audio clip's corresponding caption \(T\). We denote a minibatch of \(B\) pairs by \(\{(A_i,T_i)\}_{i=1}^B\). The MuQ encoder produces frame-level audio features \(H^a = f_{\mathrm{MuQ}}(A) \in \mathbb{R}^{B \times T \times 1024}\), where \(T = 250\) frames, and the E5 encoder produces token-level text features \(H^t = g_{\mathrm{E5}}(T) \in \mathbb{R}^{B \times L \times 1024}\), where \(L = 128\) tokens. We derive \textbf{global} embeddings by mean-pooling the audio frames and extracting the \(\mathsf{[CLS]}\) token for text as formulated below:

$$
\bar h^a_i = \frac{1}{T}\sum_{t=1}^T H^a_{i,t,:}, 
\quad
\bar h^t_i = H^t_{i,0,:}
$$

\noindent We retain the full matrices \(H^a\) and \(H^t\) for \textbf{fine-grained} learning. Both global \(\bigl(\bar h^a,\bar h^t\bigr)\) and fine-grained \(\bigl(H^a,H^t\bigr)\) features are projected into a shared 512-dimensional space via lightweight projectors. Specifically,  \(\bigl(Z^a_{\mathrm{global}},\, Z^a_{\mathrm{frame}}\bigr) \leftarrow \mathrm{AudioProj}\bigl(\bar h^a,\, H^a\bigr)\) and \(\bigl(Z^t_{\mathrm{global}},\, Z^t_{\mathrm{token}}\bigr) \leftarrow \mathrm{TextProj}\bigl(\bar h^t,\, H^t\bigr)\). Our audio and text projectors each comprise two Transformer encoder layers followed by a linear layer mapping to a 512-dimensional embedding space. After projection, we extract both global and fine-grained embeddings from the audio and text encoders. The global representations, obtained via mean-pooling, capture high-level semantic alignment across modalities, while the frame-level and token-level features preserve detailed temporal and lexical correspondences.

\begin{figure*}[t]
  \centering
  \includegraphics[width=\textwidth]{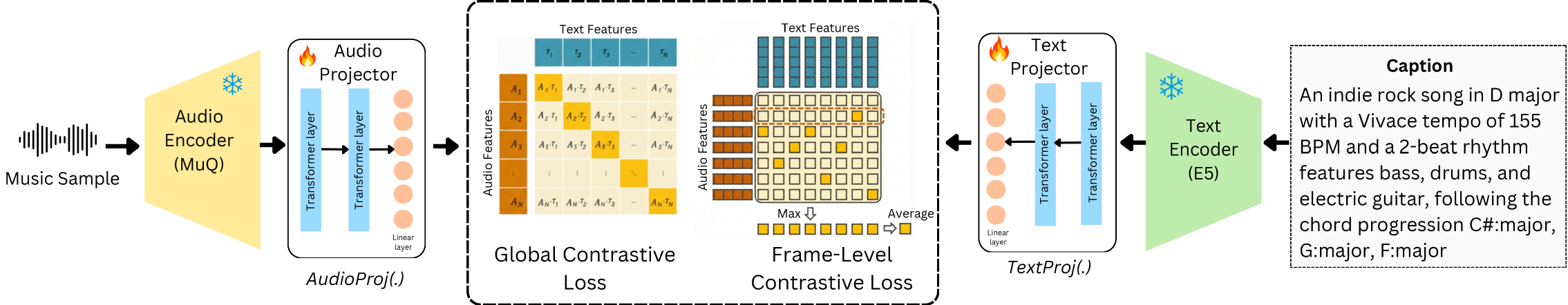}
  \caption{Architecture and training overview of FIGMA. FIGMA uses a MuQ audio encoder and E5 text encoder. It uses a weighted sum of global and frame-level local contrastive loss for training.}
  \label{fig:architecture_diagram}
\end{figure*}

\noindent We then apply our \emph{Multi-View Contrastive loss} on these embeddings, which consists of two complementary contrastive objectives: a \emph{vanilla global} contrastive loss on the global audio-text embeddings \(\{Z^a_{\mathrm{global}},Z^t_{\mathrm{global}}\}\), and a \emph{frame-level} contrastive loss on the fine-grained audio-text embeddings \(\{Z^a_{\mathrm{frame}},Z^t_{\mathrm{token}}\}\). These multi-view constraints drive the model to learn robust, semantically rich representations at both coarse and fine granularities. Detailed formulations of these losses are given in Section~\ref{subsec:gcl}, Section~\ref{subsec:fcl} and Section~\ref{subsec:mcl}, and our model's architecture is illustrated in Figure~\ref{fig:architecture_diagram}.

\subsubsection{Global Contrastive Loss}
\label{subsec:gcl}

Vanilla global contrastive learning aligns paired audio–text embeddings while repelling mismatched examples within a minibatch using the InfoNCE objective. Each matching audio–text pair \((A_i, T_i)\) is treated as a positive example, while all other pairs in the batch act as negatives. The objective encourages high similarity between matched pairs and low similarity between mismatched pairs.

\noindent Let \(\mathbf{z}^a_i = Z^a_{\mathrm{global},i}\) and \(\mathbf{z}^t_j = Z^t_{\mathrm{global},j}\) denote the global audio and text embeddings for examples \(i\) and \(j\), respectively, each normalized to unit norm. We define cosine similarity as \(\mathrm{sim}(\mathbf{u}, \mathbf{v}) = \mathbf{u}^\top \mathbf{v}\), and let \(s_{i,j} = \mathrm{sim}(\mathbf{z}^a_i, \mathbf{z}^t_j)\). The symmetric global contrastive loss for example \(i\) is then defined as:
\[
\begin{aligned}
\ell^{\mathrm{global}}_i
=
-\frac{1}{2}\Biggl(
&\log \frac{\exp\!\bigl(s_{i,i}/\tau\bigr)}
{\sum_{j=1}^{B}\exp\!\bigl(s_{i,j}/\tau\bigr)} \\
&\quad +\;
\log \frac{\exp\!\bigl(s_{i,i}/\tau\bigr)}
{\sum_{j=1}^{B}\exp\!\bigl(s_{j,i}/\tau\bigr)}
\Biggr),
\end{aligned}
\]

\noindent where \(\tau > 0\) is a temperature hyperparameter controlling the sharpness of the similarity distribution. The final global contrastive loss is obtained by averaging over the minibatch,
\(\mathcal{L}_{\mathrm{global}} = \frac{1}{B} \sum_{i=1}^{B} \ell^{\mathrm{global}}_i\). Minimizing \(\mathcal{L}_{\mathrm{global}}\) encourages matching global audio–text embeddings to align while pushing apart all other pairs. By contrasting true audio–text pairs against many in-batch negatives, the model learns to capture shared high-level semantics.

\subsubsection{Frame-Level Contrastive Loss}
\label{subsec:fcl}
To capture fine-grained temporal and lexical correspondences beyond global alignment, we introduce a frame-level contrastive loss that enforces alignment between individual audio frames and text tokens.

Let $Z^a_{\mathrm{frame},i} = [\mathbf{z}^a_{i,1}, \dots, \mathbf{z}^a_{i,T}]$ and $Z^t_{\mathrm{token},j} = [\mathbf{z}^t_{j,1}, \dots, \mathbf{z}^t_{j,L}]$ denote the $\ell_2$-normalized frame and token embeddings (each $\mathbf{z}\in\mathbb{R}^{512}$) for audio sample $i$ and caption $j$, respectively. For each audio frame $t$ in input sample $i$, we compute its maximum similarity ($s_{i,t;j}$) to any token in sample $j$:
\begin{equation}
s_{i,t;j} = \max_{1\le \ell\le L} \mathrm{sim}\bigl(\mathbf{z}^a_{i,t},\mathbf{z}^t_{j,\ell}\bigr).
\end{equation}
The max operation identifies the best-matching token for each frame, preserving fine-grained correspondences. Averaging over all frames yields the frame-level similarity score ($S_{\mathrm{frame\text{-}level}}(i,j)$) as shown below:
\begin{equation}
S_{\mathrm{frame\text{-}level}}(i,j) = \frac{1}{T}\sum_{t=1}^{T} s_{i,t;j}.
\end{equation}

We then apply InfoNCE-style contrastive losses bidirectionally by computing both audio$\to$text and text$\to$audio loss as shown below:
\begin{equation}
\ell^{a\to t, t\to a}_{i,\mathrm{frame}} = -\log \frac{\exp\bigl(S_{\mathrm{frame\text{-}level}}(i,i)/\tau\bigr)}{\sum_{j=1}^B \exp\bigl(S_{\mathrm{frame\text{-}level}}(i,j)/\tau\bigr)},
\end{equation}
The final frame-level contrastive loss averages both directions as formulated below:
\begin{equation}
\mathcal{L}_{\mathrm{frame}} = \frac{1}{2B}\sum_{i=1}^B \bigl(\ell^{a\to t}_{i,\mathrm{frame}} + \ell^{t\to a}_{i,\mathrm{frame}}\bigr).
\end{equation}
This objective treats matching audio-text pairs as positives and all others as negatives, encouraging the model to learn detailed temporal and semantic alignments that complement the global-level objective.

\begin{table*}[t]
\centering
\begin{tabular}{l@{\hspace{0.3cm}}r@{\hspace{0.3cm}}c@{\hspace{0.3cm}}c@{\hspace{0.3cm}}c@{\hspace{0.3cm}}c@{\hspace{0.3cm}}c}
\hline
\textbf{Dataset} & \textbf{\#Samples (Train/Test)} & \textbf{Chord} & \textbf{Tempo} & \textbf{Beat} & \textbf{Key} & \makecell{\textbf{Captions with music}\\\textbf{theoretic attributes}} \\
\hline
JamendoMaxCaps & 189,515 / 0 & \xmark & \xmark & \xmark & \xmark & \xmark \\
Music4All & 108,042 / 0 & \xmark & \cmark & \xmark & \cmark & \xmark \\
MTG-Jamendo & 48,709 / 2,707 & \xmark & \xmark & \xmark & \xmark & \xmark \\
MusicBench & 52,768 / 400 & \cmark & \cmark & \cmark & \cmark & \cmark \\
MusicCaps & 5,521 / 0 & \xmark & \xmark & \xmark & \xmark & \xmark \\
SDD & 1,106 / 0 & \xmark & \xmark & \xmark & \xmark & \xmark \\
\hline
FGMCaps (ours) & 380,878 / 10,000 & \cmark & \cmark & \cmark & \cmark & \cmark \\
\hline
\end{tabular}
\caption{Comparison of datasets with respect to sample count and the presence of metadata in captions.}
\label{tab:dataset-comparison}
\end{table*}

\subsubsection{Multi-View Contrastive Loss}
\label{subsec:mcl}

Our Multi-view contrastive loss is defined as:
\begingroup
\setlength{\abovedisplayskip}{6pt}
\setlength{\belowdisplayskip}{6pt}
\setlength{\abovedisplayshortskip}{2pt}
\setlength{\belowdisplayshortskip}{2pt}
$$
\mathcal{L}_{\mathrm{Multi-View}}
= \alpha\,\mathcal{L}_{\mathrm{global}}
\;+\;(1-\alpha)\,\mathcal{L}_{\mathrm{frame}},
$$
\endgroup

\noindent where \(\alpha\in[0,1]\) is a hyperparameter, which balances coarse-grained and fine-grained objectives.  Empirically, \(\alpha=0.6\) gives the best performance, leveraging high-level semantic alignment and detailed temporal correspondence.

\noindent We utilize a \textbf{multi-view contrastive loss} in which the model learns from both coarse-grained and fine-grained features by using global and frame-level contrastive losses, respectively. We use both losses rather than only the frame-level loss because the global contrastive loss helps the model capture high-level audio features and high-level semantic information in the text, with the CLS token providing a holistic representation of the caption, helping the model learn the complete caption's relation to the audio. The frame-level contrastive loss, in contrast, helps the model learn relationships between individual audio frames and specific caption tokens and is essential for capturing music-specific features and fine-grained details present in the text. As a result, both losses are complementary and necessary, and we therefore utilize both.

\subsection{FGMCaps}

\subsubsection{Dataset construction details}

To address the lack of detailed musical annotations in existing datasets, we introduce FGMCaps, comprising 380K training samples and 10K test samples with comprehensive annotations including chord progression, tempo, beat count, key, genre, and mood.

\noindent \textbf{Audio Sources.} We curate our dataset from four publicly available music collections: MTG-Jamendo \cite{mtg-jamendo}, which provides rich metadata including 55 genre tags, 87 theme tags, and 40 instrument tags; Music4All \cite{music4all} containing 108K music clips with genre metadata; JamendoMaxCaps \cite{jamendo-maxcaps} offering 200K diverse music samples; and MusicBench \cite{mustango} which provides high-quality music with detailed natural language descriptions. 

\noindent \textbf{Music Feature Extraction.} We employ state-of-the-art music theory-specific tools to extract precise musical attributes. BeatNet \cite{beatnet}, a multi-task deep neural network, extracts beat count, tempo (BPM), and time signature through particle filtering-based beat tracking. Omnizart's chord recognition module \cite{omnizart}, using a pre-trained deep learning model operating at 10 Hz frame rate, provides chord progressions with temporal boundaries. We apply post-processing to remove consecutive duplicate chords and identify repeating patterns. Essentia's KeyExtractor \cite{essentia}, implementing the Krumhansl-Schmuckler key-finding algorithm \cite{krumhansl1990cognitive} on chroma features, determines musical key, scale (major/minor), and confidence scores.

\noindent \textbf{Caption Generation.} To generate natural language descriptions incorporating extracted features and metadata, we employ Qwen3-Next-80B-A3B-Instruct \cite{qwen3technicalreport}. For each audio clip, we construct prompts containing five musical characteristic sentences derived from extracted features: BPM description, tempo text (e.g., "Allegro"), time signature, chord progression, and musical key. For MTG-Jamendo samples, we additionally include genre, theme, and instrumentation metadata. For Music4All samples, we include genre metadata. We randomize the order of musical characteristics to prevent position-dependent biases. The model is instructed to generate single-sentence captions using factual, objective language without subjective interpretations.

\begin{table*}[t]
\centering
\small
\setlength{\tabcolsep}{3pt}
\resizebox{\textwidth}{!}{%
\begin{tabular}{lcccccccc}
\toprule
 & \multicolumn{4}{c}{Text-to-Audio Retrieval} & \multicolumn{4}{c}{Audio-to-Text Retrieval}\\
\cmidrule(lr){2-5}\cmidrule(lr){6-9}
Model &
R@1 & R@5 & R@10 & R@20 &
R@1 & R@5 & R@10 & R@20\\
\midrule
LAION-CLAP$_{\text{(General Audio)}}$ &
23.86 & 48.73 & 62.94 & 77.16 &
27.92 & 53.81 & \underline{76.65} & 86.80\\
LAION-CLAP$_{\text{(General audio with variable-length)}}$ &
19.80 & 44.67 & 57.87 & 73.10 &
23.35 & 52.79 & 62.94 & 78.17\\
LAION-CLAP$_{\text{(Music)}}$ &
25.38 & 55.84 & 68.53 & 79.70 &
25.38 & 61.93 & 76.14 & \textbf{89.34}\\
LAION-CLAP$_{\text{(Music and Speech)}}$ &
19.29 & 46.19 & 64.47 & 77.66 &
19.80 & 48.22 & 65.48 & 80.20\\
LAION-CLAP$_{\text{(Music, Speech and General Audio)}}$ &
19.80 & 54.82 & 65.99 & 77.16 &
26.90 & 59.39 & 71.57 & 83.76\\
MS-CLAP 2022 &
06.09 & 17.26 & 27.41 & 41.62 &
08.63 & 22.34 & 34.52 & 49.24\\
MS-CLAP 2023 &
20.30 & 44.67 & 57.87 & 70.05 &
23.86 & 55.84 & 70.56 & 83.25\\
MuQ-MuLaN &
20.81 & 47.71 & 62.94 & 74.62 &
17.76 & 43.65 & 57.86 & 78.68\\
M2D-CLAP &
25.38 & 55.33 & 70.05 & 78.17 &
\underline{36.55} & \underline{63.96} & 75.63 & 84.77\\
CLAMP 3 &
\underline{28.43} & \underline{57.87} & \underline{74.62} & \underline{89.85} &
05.08 & 24.37 & 34.01 & 52.28\\
LAION-CLAP$_{\text{(Continued Training on FGMCaps)}}$ &
10.66 & 36.55 & 48.73 & 68.53 &
13.71 & 36.55 & 52.79 & 70.05\\
FIGMA &
\textbf{34.52} & \textbf{65.99} & \textbf{81.73} & \textbf{91.37} &
\textbf{39.09} & \textbf{68.02} & \textbf{80.71} & \underline{88.83}\\
\bottomrule
\end{tabular}}
\caption{\small Retrieval performance (R@K) on MusicBench. Best values are in \textbf{bold} and second-best values are \underline{underlined}.}
\label{tab:musicbench}
\end{table*}

\noindent \textbf{Dataset Splitting.} We employ stratified splitting to ensure balanced representation across source datasets. The test set contains 10,000 samples: 2,707 from MTG-Jamendo (all available test samples), 3,646 from Music4All, and 3,647 from JamendoMaxCaps. The validation set follows identical distribution with 10,000 samples. The training set comprises all the train-set samples from MTG-Jamendo and the remaining samples from Music4All and JamendoMaxCaps, which results in approximately 330K samples, augmented with 50K filtered music-caption pairs from MusicBench, resulting in approximately 380,000 training samples in total. For MusicBench we carefully filter the training data, retaining only samples from the AudioSet train set, ensuring no data leakage, and we utilize MusicBench's main caption as well as alternative caption to help increase caption diversity.

\noindent \textbf{Quality Control.} We implement filtering procedures to remove samples with failed feature extraction (< 0.5\% of data), validate caption completeness, and verify key extraction confidence exceeds 0.5 threshold.

\subsubsection{Comparison with other datasets}
Table 1 compares FGMCaps with existing music datasets. As shown, FGMCaps is the only dataset that contains captions with comprehensive music theory attributes like chord, tempo, beat, key at a large scale of 380K training samples and 10K test samples. MusicCaps \cite{musiccaps} provides high-quality captions but lacks music theory attributes. Music4All includes only tempo and key, whereas MTG-Jamendo provides valuable genre tags but lacks any details about chords or tempo. SDD contains only 1,106 samples. MusicBench contains music theory attributes but has a much smaller size of 52K captions as compared to FGMCaps 380K captions. With the largest scale among datasets with detailed music-theoretic annotations, FGMCaps provides a very comprehensive resource for training music retrieval models as well as their evaluation.

\section{Experiments}

\noindent \textbf{Model Architecture.} We train FIGMA using the Multi-View Contrastive loss defined in Section 3.1.4, which combines global and frame-level contrastive objectives. Our architecture employs two frozen pre-trained encoders: the MuQ audio encoder, pre-trained in a self-supervised manner on large-scale music data, and the Microsoft Multilingual E5 Large Instruct text encoder, selected for its robust multilingual language understanding. The audio encoder processes 10-second audio clips sampled at 24kHz, while the text encoder processes the corresponding captions. Both encoders remain frozen during training, totaling approximately 800M parameters, and training is restricted to the projection heads with approximately 22M trainable parameters. This design significantly reduces computational cost while enabling faster training and effective alignment.

\noindent Our audio and text projection heads each consist of two Transformer encoder layers (8 attention heads, feed-forward dimension 512) followed by a linear projection into a shared 512-dimensional embedding space. We adopt this Transformer-based design instead of linear projections due to its stronger capacity to model sequential dependencies in audio frames and text tokens, which is essential for the frame-level contrastive objective.

\noindent \textbf{Model Training.} We train FIGMA on the FGMCaps training split, consisting of 380K music-caption pairs with detailed annotations including chord progression, tempo, beat count, key, as well as genre and mood. The model is trained for 15 epochs with a batch size of 256 and utilizes early stopping. We employ the Adam optimizer with a learning rate of $1\times10^{-4}$. The temperature parameter $\tau$ in the InfoNCE loss is set to $0.07$, and the loss weighting hyperparameter $\alpha$ is set to $0.6$, balancing the global and frame-level objectives.

\begin{table*}[t]
\centering
\small
\setlength{\tabcolsep}{3pt}
\resizebox{\textwidth}{!}{%
\begin{tabular}{lcccccccc}
\toprule
 & \multicolumn{4}{c}{Text-to-Audio Retrieval} & \multicolumn{4}{c}{Audio-to-Text Retrieval}\\
\cmidrule(lr){2-5}\cmidrule(lr){6-9}
Model &
R@1 & R@5 & R@10 & R@20 &
R@1 & R@5 & R@10 & R@20\\
\midrule
LAION-CLAP$_{\text{(General Audio)}}$ &
01.90 & 07.40 & 11.30 & 18.60 &
04.00 & 12.10 & 19.10 & 27.60\\
LAION-CLAP$_{\text{(General audio with variable-length)}}$ &
00.70 & 05.10 & 09.00 & 17.40 &
02.20 & 07.30 & 11.50 & 21.10\\
LAION-CLAP$_{\text{(Music)}}$ &
02.60 & 09.40 & 14.80 & 21.60 &
03.00 & 11.60 & 18.50 & 28.10\\
LAION-CLAP$_{\text{(Music and Speech)}}$ &
02.80 & 09.50 & 14.10 & 20.20 &
02.40 & 09.80 & 17.40 & 25.50\\
LAION-CLAP$_{\text{(Music, Speech and General Audio)}}$ &
03.00 & 07.50 & 12.90 & 20.80 &
02.80 & 10.10 & 17.90 & 28.00\\
MS-CLAP 2022 &
00.60 & 03.30 & 04.60 & 08.70 &
00.80 & 03.40 & 06.80 & 12.00\\
MS-CLAP 2023 &
01.50 & 04.90 & 09.30 & 14.80 &
02.90 & 08.30 & 15.50 & 22.90\\
MuQ-MuLaN &
04.10 & 12.40 & 17.80 & 27.50 &
03.90 & 11.70 & 19.10 & 28.10\\
M2D-CLAP &
01.90 & 06.70 & 11.40 & 17.80 &
03.30 & 11.90 & 19.70 & 30.80\\
CLAMP 3 &
\underline{07.50} & \underline{20.70} & \underline{30.80} & \underline{43.10} &
01.10 & 04.10 & 06.40 & 11.70\\
LAION-CLAP$_{\text{(Continued Training on FGMCaps)}}$ &
06.10 & 18.30 & 26.50 & 36.80 &
\underline{06.00} & \underline{20.00} & \underline{30.10} & \underline{40.90}\\
FIGMA &
\textbf{13.00} & \textbf{28.00} & \textbf{37.60} & \textbf{48.60} &
\textbf{13.20} & \textbf{33.30} & \textbf{42.90} & \textbf{53.40}\\
\bottomrule
\end{tabular}}
\caption{\small Retrieval performance (R@K) on FMACaps-Eval. Best values are in \textbf{bold} and second-best values are \underline{underlined}.}
\label{tab:fmacaps_eval}
\end{table*}

\noindent \textbf{Evaluation Datasets and Metrics.} We evaluate retrieval performance using Retrieval@K for K in {1, 5, 10, 20}, measuring the percentage of queries for which the correct match appears within the top K retrieved results. We assess bidirectional retrieval: text-to-audio (T2A), where text queries retrieve audio clips, and audio-to-text (A2T), where audio queries retrieve captions. For each query, we compute cosine similarity between the query embedding and all candidates in the test set, ranking by descending similarity. We evaluate on two benchmarks: FMACaps-Eval-TestB \cite{mustango}, containing 1,000 music-caption pairs from the Free Music Archive \cite{freemusicarchive}, and MusicBench \cite{mustango}, providing high-quality music with detailed natural language descriptions. These benchmarks assess both fine-grained music retrieval capability and generalization to diverse data distributions.

\section{Results}

We evaluate FIGMA on two benchmarks: FMACaps-Eval and MusicBench. Tables 2 and 3 show that FIGMA substantially outperforms all baseline models across both text-to-audio (T2A) and audio-to-text (A2T) retrieval tasks.

\noindent \textbf{MusicBench Results.} On MusicBench (Table \ref{tab:musicbench}), FIGMA achieves 34.52\% R@1 and 65.99\% R@5 for T2A retrieval, achieving 21.4\% relative improvement over CLAMP3 in R@1. For A2T retrieval, FIGMA obtains 39.09\% R@1 and 68.02\% R@5, surpassing M2D-CLAP and all other baselines. This confirms that standard contrastive objectives are insufficient for capturing fine-grained musical attributes, indicating our multi-view objective learns more transferable representations.

\noindent \textbf{FMACaps-Eval Results.} On FMACaps-Eval (Table \ref{tab:fmacaps_eval}), FIGMA achieves 13.00\% R@1 and 28.00\% R@5 for T2A retrieval, representing a 73.3\% relative improvement over the second-best model CLAMP 3. For A2T retrieval, FIGMA achieves 13.20\% R@1 and 33.30\% R@5, more than double the performance of LAION-CLAP with continued training on FGMCaps. The strong out-of-domain performance on FMACaps-Eval demonstrates FIGMA's generalization capability.

\begin{figure}[t]
  \centering
  \includegraphics[width=\columnwidth]{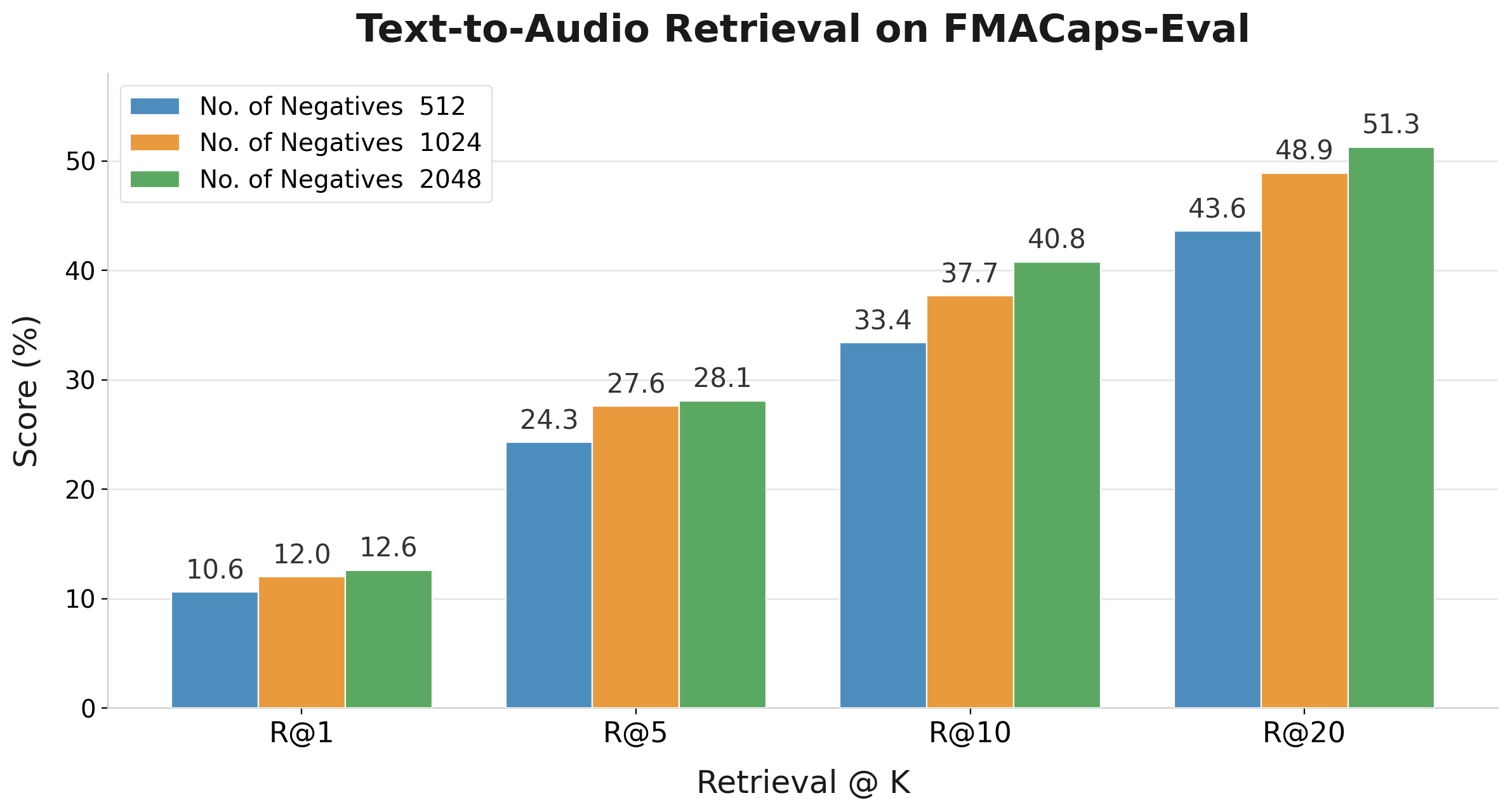}
  \caption{\small Ablation study on the effect of negative set size on different Recall@K values for FMACaps-Eval in text-to-audio retrieval.}
  \label{fig:ablation}
\end{figure}

\subsection{Negative Set Size Ablation}

We study the effect of negative set size by varying the batch size, where the number of negatives is fixed to 8× the batch size. As shown in Fig.~\ref{fig:ablation}, increasing the negative set size consistently improves recall across all K values. Larger negative pools lead to notable gains, with substantial improvements at higher recall thresholds (R@10 and R@20). This trend suggests that scaling the negative set strengthens contrastive discrimination and improves fine-grained text-to-audio retrieval performance on FMACaps-Eval.

\subsection{Robustness to Fine-Grained Attribute Perturbations}
\label{subsec:hard_negatives}

To evaluate whether FIGMA relies on genuine fine-grained alignment rather than coarse distributional cues, we construct hard-negative evaluation sets from a 3K subset of our test set. For each caption, we generate modified versions by altering exactly one attribute at a time (key, BPM, tempo marking, beat count, or chord sequence), producing five evaluation sets of 3K samples each. We then measure FIGMA's audio-to-text retrieval performance using these perturbed captions. Results are reported in Table~\ref{tab:hard_negatives}.

\noindent When a single attribute is altered, retrieval performance decreases relative to the original captions, as expected. However, FIGMA still retrieves the correct audio with strong accuracy (R@1 between 34.87\% and 43.20\%), demonstrating that the model is robust to fine-grained attribute changes rather than collapsing under small textual perturbations. This indicates that FIGMA's multi-view objective learns representations that are genuinely grounded in fine-grained musical attributes, not merely surface-level caption distributions.

\begin{table}[t]
\centering
\small
\setlength{\tabcolsep}{4pt}
\begin{tabular}{lcccc}
\toprule
\textbf{Attribute Changed} & \textbf{R@1} & \textbf{R@5} & \textbf{R@10} & \textbf{R@20} \\
\midrule
Original (no change) & 46.53 & 74.97 & 83.97 & 89.57 \\
\midrule
Key           & 38.90 & 67.97 & 77.63 & 84.77 \\
BPM           & 40.30 & 67.87 & 76.97 & 84.43 \\
Tempo marking & 35.77 & 65.87 & 75.83 & 83.97 \\
Beat count    & 34.87 & 65.17 & 74.90 & 83.57 \\
Chords        & 43.20 & 65.80 & 75.90 & 84.53 \\
\bottomrule
\end{tabular}
\caption{\small Audio-to-text retrieval performance of FIGMA on hard-negative evaluation sets. For each row, exactly one musical attribute in the caption is perturbed. FIGMA maintains strong R@1 across all perturbation types, indicating robustness to fine-grained attribute changes.}
\label{tab:hard_negatives}
\end{table}

\section{Conclusion}

We introduced FIGMA, a novel framework that advances fine-grained music retrieval by addressing a fundamental limitation of existing CLAP-based models; their inability to effectively utilize long, attribute-rich music descriptions. By combining a multi-view contrastive objective that jointly aligns global audio–text semantics and frame-level, token-wise correspondences with our large-scale FGMCaps dataset containing explicit music-theoretic annotations, FIGMA learns representations that faithfully capture both high-level musical context and precise structural attributes such as tempo, key, chord progression, and rhythm. Extensive evaluations across in-domain and out-of-domain benchmarks including MusicBench and FMACaps demonstrate that FIGMA consistently and substantially outperforms prior music retrieval models, achieving up to 73.3\% relative improvement, while remaining computationally efficient through frozen backbone encoders. 
Together, our results establish fine-grained alignment as a critical design principle for music–language models and position FIGMA as a strong foundation for music retrieval research.

\section{Limitations and Future Work}

While FIGMA achieves strong fine-grained music retrieval performance, we identify a few important limitations and future directions. First, our experiments use English-centric music captions. Although we use a multilingual text encoder, we do not explicitly evaluate cross-lingual retrieval, and future work should assess and extend FIGMA on non-English music-caption datasets to verify its multilingual generalization. Second, our training corpus: FGMCaps, focuses on tempo, key, beat count, and chord progression; future extensions could incorporate higher-order musical structure, such as song sections (e.g., verse and chorus), harmonic tension, modulation, or rhythmic motifs, to further enrich fine-grained music retrieval. Third, our frame-level objective relies on a max-operator-based similarity formulation, which does not theoretically guarantee perfectly fine-grained alignment between specific musical attributes and their corresponding acoustic features. Future work could explore softer aggregation mechanisms (e.g., attention-based or averaging strategies) to further improve attribute-level alignment.

\section*{Acknowledgements}
We would like to thank Ramaneswaran Selvakumar for his 
thoughtful feedback on the manuscript. This research is 
supported in part by Adobe, Amazon, NVIDIA, and Sesame.

\bibliography{custom}

\appendix

\section{Additional Results}
\label{app:additional_results}

\subsection{Retrieval Performance on FGMCaps Test Set}
\label{app:fgmcaps_results}

In addition to the MusicBench and FMACaps-Eval benchmarks reported in the main paper, we also evaluate all baselines and FIGMA on the FGMCaps test set, consisting of 10,000 music-caption pairs with detailed music-theoretic annotations (Section~3.2). Table~\ref{tab:fgmcaps_results} reports bidirectional retrieval performance (R@K for K $\in \{1, 5, 10, 20\}$) on this benchmark. FIGMA substantially outperforms all baselines across both retrieval directions and all values of K, with R@1 reaching 26.15\% for text-to-audio retrieval and 26.86\% for audio-to-text retrieval, compared to under 8\% for the strongest baseline. These results reinforce our main findings: standard contrastive objectives cannot effectively leverage the fine-grained musical attributes present in FGMCaps captions, whereas FIGMA's multi-view objective captures them successfully.

\begin{table*}[t]
\centering
\small
\setlength{\tabcolsep}{3pt}
\resizebox{\textwidth}{!}{%
\begin{tabular}{lcccccccc}
\toprule
 & \multicolumn{4}{c}{Text-to-Audio Retrieval} & \multicolumn{4}{c}{Audio-to-Text Retrieval}\\
\cmidrule(lr){2-5}\cmidrule(lr){6-9}
Model &
R@1 & R@5 & R@10 & R@20 &
R@1 & R@5 & R@10 & R@20\\
\midrule
LAION-CLAP$_{\text{(General Audio)}}$ &
00.09 & 00.40 & 00.68 & 01.30 &
00.24 & 00.92 & 01.55 & 02.95\\
LAION-CLAP$_{\text{(General audio with variable-length)}}$ &
00.10 & 00.46 & 00.76 & 01.43 &
00.25 & 00.82 & 01.41 & 02.19\\
LAION-CLAP$_{\text{(Music)}}$ &
00.18 & 00.62 & 01.17 & 02.18 &
00.29 & 01.43 & 02.47 & 03.94\\
LAION-CLAP$_{\text{(Music and Speech)}}$ &
00.14 & 00.58 & 01.07 & 01.94 &
00.23 & 00.90 & 01.72 & 03.08\\
LAION-CLAP$_{\text{(Music, Speech and General Audio)}}$ &
00.10 & 00.65 & 01.08 & 01.99 &
00.28 & 01.21 & 02.16 & 03.58\\
MS-CLAP 2022 &
00.08 & 00.17 & 00.32 & 00.62 &
00.06 & 00.27 & 00.45 & 00.91\\
MS-CLAP 2023 &
00.16 & 00.65 & 01.07 & 01.90 &
00.28 & 01.12 & 01.98 & 02.99\\
MuQ-MuLaN &
00.85 & 03.06 & 05.29 & 08.61 &
\underline{01.47} & \underline{04.85} & \underline{07.39} & \underline{11.16}\\
M2D-CLAP &
00.09 & 00.54 & 00.96 & 01.66 &
00.34 & 01.15 & 01.98 & 03.36\\
CLAMP 3 &
\underline{02.22} & \underline{08.16} & \underline{12.67} & \underline{18.33} &
00.35 & 01.33 & 02.32 & 04.05\\
FIGMA &
\textbf{26.15} & \textbf{52.68} & \textbf{63.64} & \textbf{74.07} &
\textbf{26.86} & \textbf{54.25} & \textbf{65.17} & \textbf{75.16}\\
\bottomrule
\end{tabular}}
\caption{\small Retrieval performance (R@K) on the FGMCaps test set. Best values are in \textbf{bold} and second-best values are \underline{underlined}.}
\label{tab:fgmcaps_results}
\end{table*}

\begin{table*}[t]
\centering
\small
\setlength{\tabcolsep}{3pt}
\resizebox{\textwidth}{!}{%
\begin{tabular}{lcccccccc}
\toprule
 & \multicolumn{4}{c}{Text-to-Audio Retrieval} & \multicolumn{4}{c}{Audio-to-Text Retrieval}\\
\cmidrule(lr){2-5}\cmidrule(lr){6-9}
Model &
R@1 & R@5 & R@10 & R@20 &
R@1 & R@5 & R@10 & R@20\\
\midrule
LAION-CLAP$_{\text{(General Audio)}}$ &
23.86 & 48.73 & 62.94 & 77.16 &
\underline{27.92} & 53.81 & \underline{76.65} & 86.80\\
LAION-CLAP$_{\text{(General audio with variable-length)}}$ &
19.80 & 44.67 & 57.87 & 73.10 &
23.35 & 52.79 & 62.94 & 78.17\\
LAION-CLAP$_{\text{(Music)}}$ &
\underline{25.38} & \underline{55.84} & \underline{68.53} & \underline{79.70} &
25.38 & \underline{61.93} & 76.14 & \textbf{89.34}\\
LAION-CLAP$_{\text{(Music and Speech)}}$ &
19.29 & 46.19 & 64.47 & 77.66 &
19.80 & 48.22 & 65.48 & 80.20\\
LAION-CLAP$_{\text{(Music, Speech and General Audio)}}$ &
19.80 & 54.82 & 65.99 & 77.16 &
26.90 & 59.39 & 71.57 & 83.76\\
LAION-CLAP$_{\text{(Continued Training on FGMCaps)}}$ &
10.66 & 36.55 & 48.73 & 68.53 &
13.71 & 36.55 & 52.79 & 70.05\\
FIGMA &
\textbf{34.52} & \textbf{65.99} & \textbf{81.73} & \textbf{91.37} &
\textbf{39.09} & \textbf{68.02} & \textbf{80.71} & \underline{88.83}\\
\bottomrule
\end{tabular}}
\caption{\small Retrieval performance (R@K) on MusicBench. Best values are in \textbf{bold} and second-best values are \underline{underlined}.}
\label{tab:musicbench_laion}
\end{table*}

\begin{table*}[t]
\centering
\small
\setlength{\tabcolsep}{3pt}
\resizebox{\textwidth}{!}{%
\begin{tabular}{lcccccccc}
\toprule
 & \multicolumn{4}{c}{Text-to-Audio Retrieval} & \multicolumn{4}{c}{Audio-to-Text Retrieval}\\
\cmidrule(lr){2-5}\cmidrule(lr){6-9}
Model &
R@1 & R@5 & R@10 & R@20 &
R@1 & R@5 & R@10 & R@20\\
\midrule
LAION-CLAP$_{\text{(General Audio)}}$ &
01.90 & 07.40 & 11.30 & 18.60 &
04.00 & 12.10 & 19.10 & 27.60\\
LAION-CLAP$_{\text{(General audio with variable-length)}}$ &
00.70 & 05.10 & 09.00 & 17.40 &
02.20 & 07.30 & 11.50 & 21.10\\
LAION-CLAP$_{\text{(Music)}}$ &
02.60 & 09.40 & 14.80 & 21.60 &
03.00 & 11.60 & 18.50 & 28.10\\
LAION-CLAP$_{\text{(Music and Speech)}}$ &
02.80 & 09.50 & 14.10 & 20.20 &
02.40 & 09.80 & 17.40 & 25.50\\
LAION-CLAP$_{\text{(Music, Speech and General Audio)}}$ &
03.00 & 07.50 & 12.90 & 20.80 &
02.80 & 10.10 & 17.90 & 28.00\\
LAION-CLAP$_{\text{(Continued Training on FGMCaps)}}$ &
\underline{06.10} & \underline{18.30} & \underline{26.50} & \underline{36.80} &
\underline{06.00} & \underline{20.00} & \underline{30.10} & \underline{40.90}\\
FIGMA &
\textbf{13.00} & \textbf{28.00} & \textbf{37.60} & \textbf{48.60} &
\textbf{13.20} & \textbf{33.30} & \textbf{42.90} & \textbf{53.40}\\
\bottomrule
\end{tabular}}
\caption{\small Retrieval performance (R@K) on FMACaps-Eval. Best values are in \textbf{bold} and second-best values are \underline{underlined}.}
\label{tab:fmacaps_eval_laion}
\end{table*}

\section{Additional Dataset Details}

We provide additional details on LAION-CLAP variants in Table~\ref{tab:musicbench_laion} and Table~\ref{tab:fmacaps_eval_laion}.

\subsection{Training Dataset Composition}

FGMCaps is constructed by aggregating multiple publicly available music datasets, each contributing complementary metadata and musical diversity. Table~\ref{tab:fgmcaps_sources} summarizes the individual source datasets and their contribution to the final training, validation, and test splits. All datasets are used strictly for research purposes and are released under licenses permitting non-commercial academic use.

We do not collect any new human annotations. All musical attributes are obtained through automatic music analysis tools, and all textual captions are generated automatically based on extracted features and existing metadata.

\begin{table}[t]
\centering
\resizebox{\columnwidth}{!}{
\begin{tabular}{lccc}
\hline
\textbf{Dataset} & \textbf{\# Train} & \textbf{\# Validation} & \textbf{\# Test} \\
\hline
MTG-Jamendo        & 48{,}709   & 2{,}707  & 2{,}707 \\
Music4All          & 100{,}750  & 3{,}646  & 3{,}646 \\
JamendoMaxCaps     & 180{,}541  & 3{,}647  & 3{,}647 \\
MusicBench         & 50{,}878 & --     & --     \\
\hline
\textbf{FGMCaps (Total)} & \textbf{380{,}878} & \textbf{10{,}000} & \textbf{10{,}000} \\
\hline
\end{tabular}
}
\caption{Source datasets used to construct FGMCaps. All datasets are publicly available and used solely for research purposes.}
\label{tab:fgmcaps_sources}
\end{table}

\subsection{Dataset Splits}

FGMCaps is split into training, validation, and test sets using stratified sampling to preserve balanced representation across source datasets. The test set consists of 10,000 samples and is disjoint from all training data. Special care is taken to avoid data leakage from MusicBench by retaining only samples originating from the AudioSet training split during FGMCaps construction.

\section{Caption Generation and Quality Control}
\label{app:caption_generation}

This section describes the automatic caption generation pipeline used to construct FGMCaps, including musical feature extraction, prompt construction, and quality control procedures. No human annotators are involved at any stage of this process.

\subsection{Automatic Musical Feature Extraction}
\label{app:feature_extraction}

To obtain fine-grained musical attributes, we apply established music analysis tools to each audio clip.

\noindent\textbf{BeatNet}: We use BeatNet, a multi-task deep neural network for joint beat, downbeat, and meter tracking, to extract tempo (in BPM), beat count, and time signature information from each audio clip. BeatNet operates using a convolutional recurrent architecture combined with particle filtering, enabling robust tempo and rhythmic structure estimation across diverse musical genres.

\noindent\textbf{Omnizart}: We extract chord progressions using the chord recognition module of Omnizart, which produces frame-level chord predictions at a fixed temporal resolution. We apply post-processing to remove consecutive duplicate chords and to identify repeating chord patterns, resulting in concise chord progression descriptions suitable for textual captioning.

\noindent\textbf{Essentia}: Musical key and scale (major or minor) are extracted using Essentia’s \texttt{KeyExtractor}, which implements a chroma-based key detection algorithm. We retain the estimated key and mode along with the associated confidence score for subsequent filtering and caption generation.

\subsection{Caption Generation Procedure}
\label{app:caption_generation_procedure}

Captions in FGMCaps are generated automatically using a large language model, \textbf{Qwen3-Next-80B-A3B-Instruct}, conditioned on structured musical attributes extracted from audio.

For each audio clip, we construct a prompt containing factual musical descriptors such as tempo, time signature, chord progression, and musical key. When available, high-level metadata provided by the source dataset (e.g., genre or instrumentation tags) is also included. The order of musical attributes in the prompt is randomized to avoid position-dependent biases.

The model is instructed to generate a single-sentence caption using objective, descriptive language. Subjective interpretations, emotional descriptions, and creative embellishments are explicitly discouraged. Each caption is generated independently and conditioned only on the extracted musical attributes and available metadata.

An example prompt used for caption generation is shown in Figure~\ref{fig:caption_prompt}.

\begin{figure}[t]
\centering
\begin{tcolorbox}[
  colback=white,
  colframe=black,
  boxrule=0.5pt,
  arc=3pt,
  left=6pt,
  right=6pt,
  top=6pt,
  bottom=6pt,
  width=\columnwidth
]
\small
\textbf{Prompt (Caption Generation).}

You are a music caption generator. Your task is to combine the following musical characteristics into a single, coherent, natural-sounding caption that describes a music audio clip.

Metadata:
- genre: dance, downtempo, electronic

Musical characteristics:
1. This song is in 4/4 time.
2. This song is in the key of F minor.
3. This track plays at 126 beats per minute.
4. This song is in Allegro.
5. This track follows the chord sequence G\#:major, C:major, G\#:major, F:minor.

Instructions:
1. Combine all the musical characteristics into a single flowing caption (ONE sentence only, can be long)
2. If metadata is provided, naturally incorporate it into the caption
3. Be factual and objective - do NOT add subjective descriptions (e.g., 'haunting', 'emotive', 'poignant', 'somber', 'vibrant')
4. Do NOT infer musical qualities not provided (e.g., 'minimal harmonic structure', 'complex arrangement', 'driving rhythm')
5. Use plain, descriptive language without emotional interpretation
6. Only output the caption text, nothing else

Caption:
\end{tcolorbox}
\caption{Example prompt used to generate captions for FGMCaps.}
\label{fig:caption_prompt}
\end{figure}

\subsection{Quality Control and Filtering}
\label{app:quality_control}

We apply multiple quality control steps to ensure the consistency and reliability of FGMCaps. Audio clips for which musical feature extraction fails are discarded. Captions that do not contain all required musical attributes or that violate formatting constraints are removed.

For musical key extraction, we retain only samples whose estimated key confidence exceeds a predefined threshold to reduce noise from uncertain predictions. We additionally verify that generated captions are non-empty and conform to the single-sentence requirement.

These filtering steps remove a small fraction of samples and ensure that all retained captions consistently describe explicit musical attributes suitable for fine-grained music retrieval.

\section{Baseline Details}

\subsection{General-Purpose Audio--Text Baselines}
\label{app:general_baselines}

We compare FIGMA against several general-purpose audio--text contrastive models that are trained on large-scale audio--caption datasets spanning diverse audio domains, including speech, environmental sounds, and music.

\noindent\textbf{LAION-CLAP}: We evaluate multiple variants of LAION-CLAP trained on different mixtures of general audio, music, and speech data. These models employ a standard global contrastive learning objective and serve as strong zero-shot baselines for cross-modal audio--text retrieval.

\noindent\textbf{MS-CLAP}: We include both the 2022 and 2023 releases of Microsoft CLAP, which are trained on millions of audio--text pairs and have demonstrated strong performance on a wide range of audio classification and retrieval tasks. Similar to LAION-CLAP, these models rely on global audio and text embeddings optimized using a contrastive objective.

These general-purpose baselines provide a reference point for evaluating fine-grained music retrieval performance and highlight the limitations of global contrastive learning when applied to long, music-specific captions.

\subsection{Music-Specific Baselines}
\label{app:music_baselines}

We additionally compare FIGMA against music-specific audio--text retrieval models that are trained exclusively or primarily on music data and are designed to capture musical structure more effectively than general-purpose leads.

\noindent\textbf{MuQ-MuLaN}: MuQ-MuLaN combines the MuQ self-supervised music audio encoder with the MuLaN contrastive framework. The model processes raw audio using a music-specialized encoder and aligns it with text using a global contrastive objective. It represents a strong baseline for music retrieval with learned music representations.

\noindent\textbf{CLAMP}: CLAMP is a contrastive language-music pretraining model that integrates symbolic music representations alongside audio features. It is designed to capture higher-level musical structure and has demonstrated competitive performance on music retrieval benchmarks.

\noindent\textbf{M2D-CLAP}: M2D-CLAP adapts CLAP-style training to music by leveraging masked audio modeling and music-domain pretraining. While it improves music representation quality, it continues to rely on global audio--text embedding alignment.

These models represent the state of the art in music-focused contrastive learning and serve as strong baselines for evaluating fine-grained music retrieval.

\section{Experimental Setup Details}

\subsection{Training Configuration}
\label{app:training_configuration}

FIGMA is trained on the FGMCaps training split using the multi-view contrastive loss described in Section~3.1.4. Training is conducted for a fixed number of epochs with early stopping based on validation performance.

We use the Adam optimizer with a learning rate of $1 \times 10^{-4}$ and a batch size of 256. The temperature parameter $\tau$ in the contrastive loss is set to 0.07, and the loss weighting parameter $\alpha$ balancing the global and frame-level objectives is set to 0.6. All other hyperparameters are kept consistent across experiments.

\subsection{Software}
\label{app:software}

All models are implemented and trained using PyTorch. We use DeepSpeed to enable efficient multi-GPU training and to support large batch sizes during contrastive learning. Pre-trained audio and text encoders are loaded using the Hugging Face Transformers library, which also provides tokenization and model configuration utilities.

Unless otherwise specified, all experiments use standard library implementations without custom CUDA kernels.

\subsection{Model Size and Compute}
\label{app:model_size_compute}

FIGMA employs two frozen pre-trained encoders: the MuQ audio encoder and the Multilingual E5 Large Instruct text encoder. Together, these encoders comprise approximately 800M parameters and remain fixed during training. Only the audio and text projection heads are trained, resulting in approximately 22M trainable parameters.

By freezing the encoders and training lightweight projection heads, FIGMA significantly reduces computational cost compared to end-to-end fine-tuning of large audio--text models. This design enables efficient training while preserving strong representation quality for fine-grained music retrieval.

All experiments are conducted using 8 NVIDIA A100 GPUs. Multi-GPU training is enabled via DeepSpeed to support large batch sizes and efficient contrastive learning.

\section{AI Assistants Usage}
\label{app:ai_usage}

AI assistants are used in a limited and clearly defined manner in this work.

First, a large language model is used for \emph{data generation} during the construction of the FGMCaps dataset. Specifically, an LLM is employed to generate natural language captions from automatically extracted musical attributes such as tempo, key, chord progression, and beat structure, as described in Appendix~B. The model is instructed to produce factual, single-sentence descriptions without subjective interpretation. No human annotators are involved in this process.

Second, AI-based writing assistants are used during manuscript preparation for \emph{grammar checking, language polishing, and improving clarity}. These tools do not contribute to the scientific content, experimental design, analysis, or conclusions of the paper.

\section{Potential Risk}

The model could be misused to more efficiently search or organize copyrighted music collections, potentially enabling unlicensed reuse or large-scale dataset repurposing. Such risks arise from downstream use rather than the intended research application.

\end{document}